\documentclass[conference]{IEEEtran}
\usepackage{mathrsfs}
\usepackage[square,numbers,sort&compress]{natbib}

\usepackage{ifpdf}

\usepackage{amsmath}
\usepackage{amssymb}
\usepackage{enumerate}
\usepackage{color,xcolor}
\usepackage{graphicx}
\usepackage{subfigure}
\usepackage{pifont}

\ifCLASSINFOpdf
\else
\fi

\newtheorem{algorithm}{\textbf{Algorithm}}
\begin{document}
%

\title{A New Ensemble of Rate-Compatible LDPC Codes}



\author{\authorblockN{Kai Zhang\authorrefmark{1}, Xiao Ma\authorrefmark{1}, Shancheng Zhao\authorrefmark{1}, Baoming Bai\authorrefmark{2} and Xiaoyi Zhang\authorrefmark{3}
\authorblockA{\authorrefmark{1}Department of Electronics and Communication
Engineering, Sun Yat-sen University, Guangzhou 510006, GD, China}
\authorblockA{\authorrefmark{2}State Lab. of ISN, Xidian University, Xi'an 710071, Shaanxi, China}
\authorblockA{\authorrefmark{3}National Digital Switching System Engineering and Technological R\&D Center, Zhengzhou 450002, China}}
Email: maxiao@mail.sysu.edu.cn}


%


\maketitle

\begin{abstract}
 In this paper, we presented three approaches to improve the design of Kite codes~(newly proposed rateless codes), resulting in an ensemble of rate-compatible
 LDPC codes with code rates varying ``continuously" from 0.1 to 0.9 for additive white Gaussian noise~(AWGN) channels. The new ensemble rate-compatible LDPC codes can be constructed conveniently with an empirical formula. Simulation results show that, when applied to incremental redundancy hybrid automatic repeat request~(IR-HARQ) system, the constructed codes~(with higher order modulation) perform well in a wide range of signal-to-noise-ratios~(SNRs).
\end{abstract}




%
\IEEEpeerreviewmaketitle

\section{Introduction}
Rate-compatible low-density parity-check~(RC-LDPC) codes, which may find applications in hybrid automatic repeat request~(HARQ) systems, can be constructed in at least two ways. One way is to puncture~(or extend) properly a well-designed mother code~\cite{Ha04, Yazdani04, Hsu08, El-Khamy09, Kim09, Vellambi09}. Another way is to turn to rateless coding~\cite{Luby02, Shokrollahi06, Etesami06, Palanki04}, resulting in rate-compatible codes with incremental redundancies~\cite{Soijanin06}.

Recently, the authors proposed a new class of rateless codes, called Kite codes~\cite{Ma11}, which are a special class of prefix rateless low-density parity-check~(LDPC) codes. Kite codes have the following nice properties. First, the design of Kite codes can be conducted progressively using a one-dimensional optimization algorithm. Second, the maximum-likelihood decoding performance of Kite codes with binary-phase-shift-keying~(BPSK) modulation over AWGN channels can be analyzed. Third, the definition of Kite codes can be easily generalized to groups~\cite{Ma11b}. In this paper, we attempt to improve the design of Kite codes and investigate their performance when combined with high-order modulations.

The rest of this paper is organized as follows. In Section~\ref{sec:Review of Kite Codes}, we review the construction of the original Kite codes. The design of Kite codes and existing issues are discussed. In Section~\ref{sec:Improved Design of Kite Codes}, we present three approaches to improve the design of Kite codes, resulting in good rate-compatible codes with rates varying continuously from 0.1 to 0.9 for additive white Gaussian noise~(AWGN) channels. In section~\ref{sec:Applications of Improved Kite codes}, we present simulation results for the application of Kite codes to the  HARQ system. Section~\ref{sec:Conclusion} concludes this paper.

\section{Review of Kite Codes}\label{sec:Review of Kite Codes}

\subsection{Definition of Kite Codes}
An ensemble of Kite codes, denoted by $\mathcal{K}[\infty,k;\underline{p}]$, is specified by its dimension $k$ and a so-called {\em p-sequence} $\underline p = (p_0, p_1, \cdots, p_{t}, \cdots)$ with $0< p_t < 1$ for $t\geq 0$.
A codeword $\underline c = (\underline v, \underline w) \in \mathcal{K}[\infty,k;\underline{p}]$ consists of an information vector $\underline v = (v_0, v_1, \cdots, v_{k-1})$ and a parity sequence $\underline w = (w_0, w_1, \cdots, w_t, \cdots)$. The parity bit at time $t\geq 0$ can be computed recursively by $w_t = w_{t-1} + \sum_{0\leq i < k} h_{t, i}v_i$ with $w_{-1} = 0$, where $h_{t, i}$~($0\leq i  < k$) are $k$ independent realizations of a Bernoulli distributed variable with success probability $p_t$.

For convenience, the prefix code with length $n$ of a Kite code is also called a Kite code and simply denoted by $\mathcal{K}[n, k]$. A Kite code $\mathcal{K}[n,k]$ for $n\geq k$ is a systematic linear code with $r\stackrel{\Delta}{=}n-k$ parity bits. We can write its parity-check matrix $H$ as
\begin{equation}\label{parity-check-matrix}
       H = \left(H_v, H_w\right),
\end{equation}
where $H_v$ is a matrix of size $r\times k$ that corresponds to the information bits, and $H_w$ is a square matrix of size $r\times r$ that corresponds to the parity bits. By construction, we can see that the sub-matrix $H_v$ is a random-like binary matrix whose entries are governed by the {\it p}-sequence. More specifically,
\begin{equation}\label{binaryvariable}
            {\rm Pr}\{H_{t, i} = h_{t, i}\} = \left\{\begin{array}{cc}
              p_t, & h_{t,i} = 1 \\
              1-p_t, & h_{t,i} = 0 \\
            \end{array} \right.
        \end{equation}
for $t\geq 0$ and $0\leq i < k$. In contrast, the square matrix $H_w$ is a dual-diagonal matrix~(blanks represent zeros)
\begin{equation}\label{parity-check-matrix-right}
    H_w = \left(\begin{array}{ccccc}
      1         &           &          &           &        \\
      1         & 1         &          &           &        \\
                & 1         & \ddots   &           &        \\
                &           & \ddots   & 1         &        \\
                &           &          & 1         &1       \\
    \end{array}\right).
\end{equation}

In the case of $p_t \ll 1/2$, a Kite code can be decoded as an LDPC code.


\subsection{Relationships between Kite Codes and Existing Codes}
A specific realization of a Kite code with finite length can be considered as a serially concatenated
code with a systematic low-density generator-matrix~(LDGM)
code~\cite{Garcia-Frias03} as the outer code and an accumulator as
the inner code. Different from conventional serially concatenated
codes, the inner code takes only the parity bits from the
outer code as input bits. A specific Kite code is also similar to the
generalized irregular repeat-accumulate~(GeIRA) codes~\cite{Jin00}\cite{Yang04}\cite{Liva05}\cite{Abbasfar07}.

As an ensemble, Kite codes are new. To clarify this, we resort to the definition of a code ensemble~\cite{Gallager68}.
A binary linear code {\em ensemble} is a probability space $(\mathcal{C}, Q)$, which is specified by a sample space $\mathcal{C}$ and a probability assignment $Q(\mathbf{C})$ to each $\mathbf{C} \in \mathcal{C}$.  Each sample $\mathbf{C} \in \mathcal{C}$ is a binary linear code, and the probability $Q(\mathcal{C})$ is usually implicitly determined by a random construction method. The following examples show us three code ensembles with length $n$.
\begin{itemize}
  \item A code ensemble $\mathcal{C}_g$ can be characterized by a generator matrix $G$ of size $k\times n$, where each element of $G$ is drawn independently from a uniformly distributed binary random variable.
  \item A code ensemble $\mathcal{C}_h$ can be characterized by a parity-check matrix $H$ of size $(n-k)\times n$, where each element of $H$ is drawn independently from a uniformly  distributed binary random variable.
  \item A code ensemble $\mathcal{C}_s$ can be characterized by a generator matrix $G_s = [I_k, P]$ of size $k\times n$, where $I_k$ is the identity matrix and each element of $P$ is drawn independently from a uniformly distributed binary random variable. This ensemble can also be characterized by a random parity-check matrix $H_s = [P^T, I_{n-k}]$.
\end{itemize}

In a strict sense, the above three ensembles are different. The ensemble $\mathcal{C}_g$ has some samples with code rates less than $k/n$, and the ensemble $\mathcal{C}_h$ has some samples with code rates greater than $k/n$. The code ensemble $\mathcal{C}_s$ is systematic and each sample has the exact code rate $k/n$. LT-codes~(prefix codes with length $n$) have a slightly different sample space from that of $\mathcal{C}_g$ but, with high probability, have low-density generator matrices. In contrast, the ensemble of Kite codes~(of length $n$) with $p_t \equiv 1/2$ is the same as $\mathcal{C}_s$. An ensemble of Kite codes~(of length $n$) with $p_t  < 1/2$ has the same sample space as that of $\mathcal{C}_s$ but different probability assignment to each code.

\subsection{Design of Kite Codes}
Given a data length $k$. The task to optimize a Kite code is to select the whole $p$-sequence such that all the prefix codes are good enough. This is a multi-objective optimization problem and could be very complex. For simplicity, we consider only codes with rates greater than $0.05$ and partition equally the code rates into $19$ subintervals, that is,

\begin{equation}
       (0.05, 1] = \bigcup_{1\leq \ell \leq 19} (0.05\ell, 0.05(\ell+1)].
\end{equation}
We simply assume that the $p$-sequence is a step function of the code rates taking only 19 possibly different values $\{q_\ell, 1\leq \ell \leq 19\}$. That is, we enforce that $p_t = q_\ell$ whenever $t$ satisfies that the code rate $k/(t+k) \in (0.05\ell, 0.05(\ell+1)]$. Then our task is transformed into selecting the parameters $\{q_\ell, 1\leq \ell \leq 19\}$. One approach to do this is the following greedy optimization algorithm.

First, we choose $q_{19}$ such that the prefix code of length $\lfloor k/0.95 \rfloor$ is as good as possible. Secondly, we choose $q_{18}$ with fixed $q_{19}$ such that the prefix code of length $\lfloor k/0.90 \rfloor$ is as good as possible. Thirdly, we choose $q_{17}$ with fixed $(q_{19}, q_{18})$ such that the prefix code of length $\lfloor k/0.85 \rfloor$ is as good as possible. This process continues until $q_1$ is selected.

Let $q_{\ell}$ be the parameter to be optimized. Since the parameters $q_j$ with $j > \ell$ have been fixed and the parameters $q_j$ with $j < \ell$ are irrelevant to the current prefix code, the problem to design the current prefix code then becomes a one-dimensional optimization problem, which can be solved, for example, by the golden search method~\cite{Rardin98}. What we need to do is to make a choice between any two candidate parameters $q_{\ell}$ and $q'_{\ell}$. This can be done with the help of simulation, as illustrated in~\cite{Ma11}.

\subsection{Existing Issues}
The above greedy optimization algorithm has been implemented in~\cite{Ma11}, resulting in good Kite codes. However, we have also noticed the following issues.
\begin{enumerate}
  \item In the high-rate region, Kite codes suffer from error-floors at BER around $10^{-4}$.
 
  \item In the low-rate region, there exists a relatively large gap between the performances of the Kite codes and the Shannon limits.
 
  \item The optimized {\it p}-sequence depends on the data length $k$ and has no closed form. Therefore, we have to search the {\it p}-sequence for different data lengths when required. This inconvenience definitely limits the applications of the Kite codes.
\end{enumerate}

The first issue has been partially solved by either taking RS codes as outer codes in~\cite{Ma11} or inserting fixed patterns into the parity-check matrix in~\cite{Bai11}. The objective of this paper is to provide simple ways to overcome the above issues.

\section{Improved Design of Kite Codes}\label{sec:Improved Design of Kite Codes}
As mentioned in the preceding section, we consider only code rates greater than $0.05$ and partition equally the interval $(0.05, 1]$ into 19 sub-intervals. Given a data length $k$. Define $n_{\ell} = \lfloor \frac{k}{0.05 \ell}\rfloor$ for $1\leq \ell \leq 20$.

\subsection{Constructions of the Parity-Check Matrix}
We need to construct a parity-check matrix $H=(H_v, H_w)$ of size $(n_1-k)\times n_1$, where $H_v$ is a matrix of size $(n_1-k)\times k$ corresponding to the information bits and $H_w$ is a lower triangular matrix with non-zero diagonal entries. Let $\mathcal{C}[n_1, k]$ be the code defined by $H$. We then have a sequence of prefix codes with incremental parity bits. Equivalently, we can have a sequence of prefix codes with rates varying ``continuously" from $0.05$ to 1.

To describe the algorithm more clearly, we introduce some definitions. Let $H^{(\ell)} = (H_v^{(\ell)}, H_w^{(\ell)})$ be the parity-check matrix for the prefix code with length $n_{\ell}$. Let $H_v^{(\ell)}(\cdot,j)$ and $H_v^{(\ell)}(t,\cdot)$ be the $j$-th column and $t$-th row of $H_v^{(\ell)}$, respectively. Sometimes, we use
$H_v^{(\ell)}(t,j)$ to represent the entry of $H_v^{(\ell)}$ at the location $(t,j)$.
Given $\{q_{\ell}, 1\leq \ell \leq 19\}$, the parity-check matrix $H = H^{(1)}$ can be constructed as follows.

First, the matrix $H_v$ corresponding to the information bits can be constructed progressively as shown in the following algorithm.

\begin{algorithm}{(Row-weight Concentration)}\label{Row-weight Concentration}
    \begin{enumerate}
          \item Initially, set $H_v^{(20)}$ be the empty matrix and $\ell = 19$.

          \item While $\ell \geq 1$, do the following.

          \begin{enumerate}
                \item {\em Initialization}:  generate a random binary matrix $H_v^{(\delta)}$ of size $(n_{\ell}-n_{\ell+1})\times k$, where each entry is independently drawn from a Bernoulli distribution with success probability $q_{\ell}$; form the matrix corresponding to the information bits as
                    \begin{equation}\label{ProgressiveHv}
                        H_v^{(\ell)} = \left(
                            \begin{array}{c}
                              H_v^{(\ell+1)} \\
                              H_v^{(\delta)} \\
                            \end{array}
                            \right).
                    \end{equation}

                \item {\em Row-weight concentration:}
                     \begin{enumerate}
                          \item Find $t_1~(n_{\ell+1}-k \leq t_1 < n_{\ell}-k)$ such that the row $H_v^{(\ell)}(t_1,\cdot)$ has the maximum Hamming weight, denoted by $W_{\max}$;
                          \item Find $t_0~(n_{\ell+1}-k \leq t_0 < n_{\ell}-k)$ such that the row $H_v^{(\ell)}(t_0,\cdot)$ has the minimum Hamming weight, denoted by $W_{\min}$;
                          \item If $W_{\max} = W_{\min}$ or $W_{\max} = W_{\min}+1$, set $\ell \leftarrow \ell-1$ and go to Step 2);
                              otherwise, go to the next step;
                          \item Find $j_1~(0 \leq j_1 \leq k-1)$ such that $H_v^{(\ell)}(t_1, j_1) = 1$ and that the column $H_v^{(\ell)}(\cdot, j_1)$ has the maximum Hamming weight;
                          \item Find $j_0~(0 \leq j_0 \leq k-1)$ such that $H_v^{(\ell)}(t_0, j_0) = 0$ and that the column $H_v^{(\ell)}(\cdot, j_0)$ has the minimum Hamming weight;
                          \item Swap $H_v^{(\ell)}(t_0, j_0)$ and $H_v^{(\ell)}(t_1, j_1)$; go to Step i).
                    \end{enumerate}

        \end{enumerate}

    \end{enumerate}
\end{algorithm}

{\bf Remarks.} From the above algorithm, we can see that the incremental sub-matrix $H_v^{(\delta)}$  corresponding to the information vector is finally modified into a sub-matrix with rows of weight $W_{\min}$ or $W_{\min}+1$. Such a modification is motivated by a theorem as shown in~\cite{Chung01a} stating that the optimal degree for check nodes can be selected as a concentrated distribution. Such a modification also excludes columns with extremely low weight.

Second, the matrix $H_w$ corresponding to the parity bits is constructed as a random accumulator specified in the following algorithm.

\begin{algorithm}{(Accumulator Randomization)}\label{Accumulator Randomization}
     \begin{enumerate}
          \item Initially, $H_w$ is set to be the identity matrix of size $(n_1-k)\times (n_1-k)$.
          \item For $t = 0, 1, \cdots, n_1-k-2$, do the following step by step.
                \begin{enumerate}
                       \item Find the maximum integer $T$ such that the code rates $k/(k + T)$ and $k/(k + t + 1)$ falls into the same subinterval, say $(0.05\ell, 0.05(\ell+1)]$;
                       \item Choose uniformly at random an integer $i_1 \in [t+1, T]$;
                       \item Set the $i_1$-th component of the $t$-th column of $H_w$ to be $1$, that is, set $H_w(i_1, t) = 1$.
               \end{enumerate}
     \end{enumerate}
\end{algorithm}

{\bf Remarks.} The accumulator randomization approach introduces more randomness to the code such that the current parity bit depends randomly on previous parity bits. We also note that, for all $\ell$, $H_w^{(\ell)}$ has the property that each of all columns but the last one has column weight 2. It is worth pointing out that both the row-weight concentration algorithm and the accumulator randomization algorithm are executed in an off-line manner. To construct the prefix code of length $n_{\ell}$, both of these two algorithms modify only the incremental $n_{\ell} - n_{\ell+1}$ rows of the parity-check matrix associated with the original Kite code, which do not affect other rows.

\subsection{Greedy Optimization Algorithms}
\begin{figure}
\centering
\includegraphics[width=7.0cm]{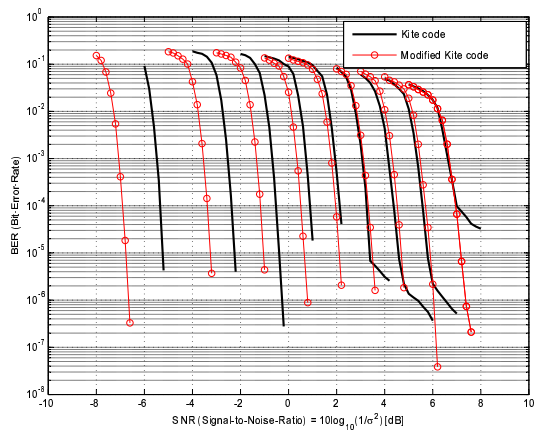}
\caption{Performances of the constructed improved Kite code and Kite code with $k = 1890$. The maximum iteration number is 50. From left to right, the curves correspond to rates 0.1, 0.2, 0.3, 0.4, 0.5, 0.6, 0.7, 0.8, and 0.9, respectively.} \label{Kite and MKite 1890}
\end{figure}

\begin{figure}
\centering
\includegraphics[width=7.0cm]{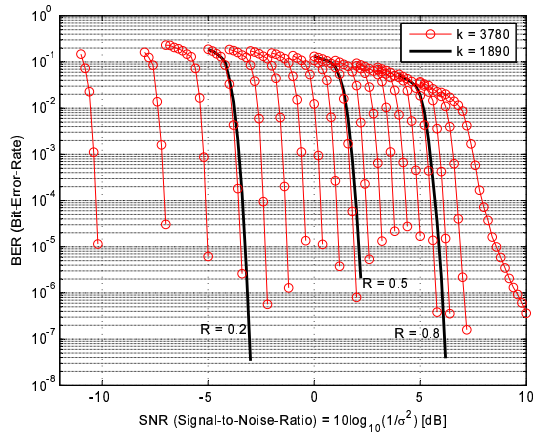}
\caption{Performance of the constructed improved Kite code with $k = 3780$. The maximum iteration number is 50. From left to right, the curves correspond to rates 0.05, 0.10, 0.15, $\cdots$, 0.90, and 0.95, respectively.} \label{MKite_code_3780}
\end{figure}

It has been shown that, given $\{q_{\ell}, 1\leq \ell \leq 19\}$, we can construct a parity-check matrix $H$ by conducting the row-weight concentration Algorithm~\ref{Row-weight Concentration} and the accumulator randomization Algorithm~\ref{Accumulator Randomization}. Hence, we can use the greedy optimization algorithm to construct a good parity-check matrix. We have designed two improved Kite codes with data length $k = 1890$ and $k = 3780$. The $p$-sequences are shown in Table~\ref{p sequence value}. Their performances are shown in Fig.~\ref{Kite and MKite 1890} and Fig.~\ref{MKite_code_3780}, respectively.

\begin{table}
\tabcolsep 2mm \caption{The $p$-sequences}\label{p sequence value}
\begin{center}

\begin{tabular}{|c|c|c|}

   \cline{1-3}
   \textrm{Code rate} $k/(k + t)$  & $p_t (k=1890)$ &  $p_t (k=3780)$  \\ \cline{1-3}

  (0.95, 1.00] &  0.0380   &   0.0170  \\    \cline{1-3}
  (0.90, 0.95] &  0.0200   &   0.0110  \\    \cline{1-3}
  (0.85, 0.90] &  0.0130   &   0.0050  \\    \cline{1-3}
  (0.80, 0.85] &  0.0072   &   0.0039  \\    \cline{1-3}
  (0.75, 0.80] &  0.0046   &   0.0023  \\    \cline{1-3}
  (0.70, 0.75] &  0.0038   &   0.0020  \\    \cline{1-3}
  (0.65, 0.70] &  0.0030   &   0.0016  \\    \cline{1-3}
  (0.60, 0.65] &  0.0028   &   0.0013  \\    \cline{1-3}
  (0.55, 0.60] &  0.0018   &   0.0010  \\    \cline{1-3}
  (0.50, 0.55] &  0.0017   &   0.0009  \\    \cline{1-3}
  (0.45, 0.50] &  0.0015   &   0.0007  \\    \cline{1-3}
  (0.40, 0.45] &  0.0014   &   0.0007  \\    \cline{1-3}
  (0.35, 0.40] &  0.0013   &   0.0006  \\    \cline{1-3}
  (0.30, 0.35] &  0.0012   &   0.0006  \\    \cline{1-3}
  (0.25, 0.30] &  0.0012   &   0.0005  \\    \cline{1-3}
  (0.20, 0.25] &  0.0012   &   0.0005  \\    \cline{1-3}
  (0.15, 0.20] &  0.0011   &   0.0005  \\    \cline{1-3}
  (0.10, 0.15] &  0.0011   &   0.0004  \\    \cline{1-3}
  (0.05, 0.10] &  0.0011   &   0.0004  \\    \cline{1-3}
\end{tabular}
\end{center}
\end{table}

From Fig.~\ref{Kite and MKite 1890} and Fig.~\ref{MKite_code_3780}, we have the following observations.
\begin{itemize}
  \item The improved Kite codes perform well within a wide range of SNRs.
  \item In the moderate-to-high-rate region, the improved Kite codes perform almost as well as the original Kite codes in the water-fall region, while the improved Kite codes can effectively lower down the error-floors.
  \item In the low-rate region, the improved Kite codes are better than the original Kite codes. For example, when the code rate is 0.1, the improved Kite code can achieve a performance gain about 1.5 dB over the original Kite code at $\textrm{BER} = 10^{-5}$.
  \item As the data length $k$ increases, the performance of the improved Kite codes can be improved. For instance, when the code rate is 0.2, the improved Kite code with data length 3780 achieves a performance gain about 0.2 dB over the improved Kite code with data length 1890 around $\textrm{BER} = 10^{-5}$.
\end{itemize}

\begin{figure}
\centering
\includegraphics[width=7.0cm]{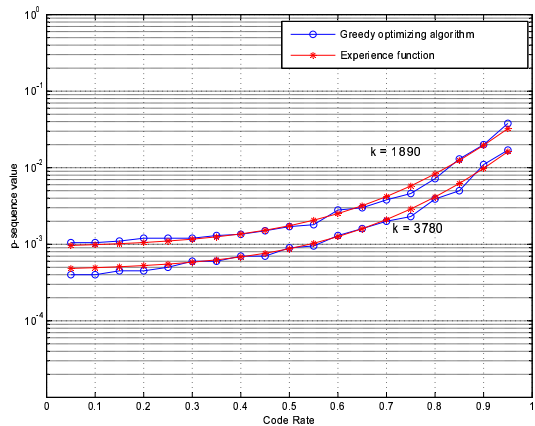}
\caption{Comparisons of the empirical formula for $q_{\ell}$ with the parameters obtained by the simulation-based greedy optimizing algorithm.
} \label{experience_function}
\end{figure}

\subsection{Empirical Formula}
As shown by the above two examples, the optimized {\it p}-sequence depends on data length $k$, which makes the design of improved Kite codes time-consuming. To accelerate the design of improved Kite codes, we present an empirical formula for the {\it p}-sequence. To this end, we have plotted the optimized parameters in Table~\ref{p sequence value} in Fig.~\ref{experience_function}. We then have the following empirical formula
\begin{equation}\label{p squence formula}
            q_{\ell} = \frac{1}{k} \left(\frac{1.65}{(1.5 - 0.05\ell)^6} + 2.0 \right)
\end{equation}
for $1\leq \ell \leq 19$. From Fig.~\ref{experience_function}, we can see that the above formula is well matched to the optimized {\it p}-sequence.

\section{Applications of Kite codes to IR-HARQ Systems}\label{sec:Applications of Improved Kite codes}
In summary, for an arbitrarily given data length $k$, we can generate~(off-line) a systematic mother code $\mathcal{K}[n_1, k]$ with $n_1 = \lfloor k / 0.05\rfloor$ by constructing a parity-check matrix according to the following procedure.
\begin{enumerate}
  \item Calculate the $q$-sequence according to~(\ref{p squence formula});
  \item Generate a semi-random parity-check matrix $H = (H_v, H_w)$ according to~(\ref{binaryvariable}) and (\ref{parity-check-matrix-right});
  \item Modify $H$ by performing Algorithm~\ref{Row-weight Concentration} and Algorithm~\ref{Accumulator Randomization}.
\end{enumerate}

We consider to combine the Kite code with two-dimensional modulations. The coded bits are assumed to be modulated into a constellation of size $2^b$ by the (near-)Gray mapping and then transmitted over AWGN channels. The receiver first extracts the {\em a posteriori} probabilities of coded bits from each noisy coded symbol and then performs the iterative sum-product decoding algorithm. If the receiver can decode successfully, it feedbacks an ``ACK"; otherwise, the transmitter produces more parity bits and transmits more coded signals. The receiver tries the decoder again with the incremental noisy coded signals. This process continues until the decoding is successful. To guarantee with high probability that the successfully decoded codeword is the transmitted one, we require that the receiver starts the decoding only after it receives $n_{18}$ noisy coded bits. With this constraint, we have never found erroneously decoded codewords in our simulations. That is, the decoder either delivers the transmitted codeword or reports a decoding failure. If $n$ is the length of the code at which the decoding is successful, we call $\eta = k b / n$ bits/symbol the decoding spectral efficiency, which is a random variable and may vary from frame to frame.

\begin{figure}
\centering
\includegraphics[width=7.0cm]{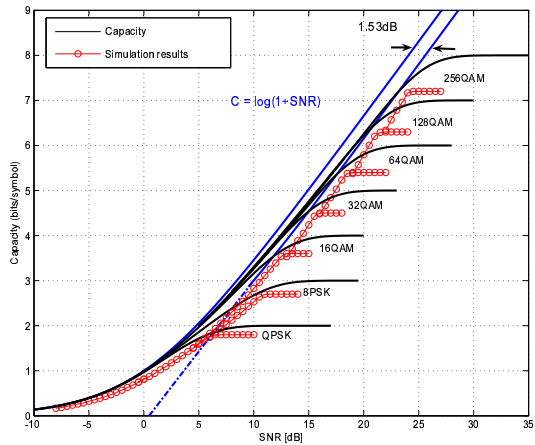}
\caption{The average decoding spectral efficiency~(at ``zero" error probability) of the improved Kite code with data length $k = 9450$ over AWGN channels. The maximum iteration number is 50.} \label{decoding_rate_BICM}
\end{figure}

The average decoding spectral efficiency of the Kite code of dimension $k = 9450$ with different high-order modulations for AWGN channels are shown in Fig.~\ref{decoding_rate_BICM}. Also shown are the capacity curves. Taking into account the performance loss of shaping gain $1.53$ dB caused by conventional QAM constellations for large SNRs~\cite{Forney84}, we conclude that the improved Kite codes perform well essentially in the whole SNR range.

\section{Conclusion and Future Works}\label{sec:Conclusion}
In this paper, we have presented three approaches to improve the design of Kite codes. In the moderate-to-high-rate region, the improved Kite codes perform almost as well as the original Kite codes, but exhibit lower error-floors. In the low-rate region, the improved Kite codes are better than the original ones. In addition, we have presented an empirical formula for the $p$-sequence, which can be utilized to construct rate-compatible codes of any code rates with arbitrarily given data length. Simulation results show that, when combined with high-order modulations, the improved Kite codes perform well essentially in the whole range of SNRs.

Future works include to compare Kite codes with other codes such as Raptor codes in terms of both the performance and the complexity when applied to adaptive coded modulations over noisy channels.

\section*{Acknowledgment}
This work was supported by the 973 Program~(No.2012CB316100) and the NSF under Grants 61172082 and 60972046 of China.

\ifCLASSOPTIONcaptionsoff
\newpage
\fi
\bibliographystyle{IEEEtran}
\bibliography{tzzt}

\end{document}